\documentstyle[aps,twocolumn]{revtex}
\begin{document}
\title{Scheme for preparation of nonmaximal entanglement between two atomic
ensembles }
\author{Peng Xue\thanks{%
Email address: xuepeng@mail.ustc.edu.cn}, and Guang-Can Guo\thanks{%
Email address: gcguo@ustc.edu.cn}}
\address{Key Laboratory of Quantum Information, University of Science and Technology\\
of China, Hefei 230026, P. R. China}
\maketitle

\begin{abstract}
\baselineskip12ptWe propose an experimentally feasible scheme to generate
nonmaximal entanglement between two atomic ensembles. The degree of
entanglement is readily tunable. The scheme involves laser manipulation of
atomic ensembles, adjustable quarter- and half-wave plates, beam splitter,
polarizing beam splitters, and single-photon detectors, and well fits the
status of the current experimental technology. Finally we use the
nonmaximally entangled state of ensembles to demonstrate quantum nonlocality
by detecting the Clauser-Horne-Shimony-Holt inequality.

PACS number(s): 03.65.Ud, 03.67.-a, 42.50.Gy, 42.50.-p
\end{abstract}

\baselineskip12ptQuantum entanglement is one of the most striking features
of quantum mechanics. The recent surge of interest and progress in quantum
information theory allows one to take a more positive view of entanglement
and regard it as an essential resource for many ingenious applications such
as quantum computation \cite{Shor,Grover,Ekert1}, quantum teleportation \cite
{Bennett1,Pan1}, superdense coding \cite{Bennett3}, and quantum cryptography 
\cite{Ekert2,Lo,xue1,xue2}. The technology of generation and manipulation of
bipartite or multipartite entangled states has been realized in some systems 
\cite{Kwiat,Kwiat1,Kwiat2,other,Bou,Rau,Sac,Pan2}. In most of the above
schemes, the subsystems are taken as single-particle. Remarkably, Lukin and
Duan {\it et al. }have proposed some schemes \cite{Lukin,Duan1,Duan2,Duan3}
for preparation of entanglement, in which atomic ensembles with a large
number of identical atoms are used as the basic system. For example, one can
use atomic ensembles for generation of substantial spin squeezing \cite{spin}
and continuous variable entanglement \cite{Duan1,Jul}, and for efficient
preparation of Einstein-Podolsky-Rosen (EPR) \cite{Duan2},
Greenberger-Horne-Zeilinger (GHZ) type of maximally entangled states \cite
{Duan3} and W class of maximally entangled states \cite{xue3}. The schemes
have some special advantages compared with other quantum information schemes
based on the control of single particles \cite{Duan4}.

In all experimental efforts, it is hard to vary the degree of entanglement,
and to produce nonmaximally entangled states without compromising the purity
of the state. Nonmaximally entangled states have been shown to reduce the
required detector efficiencies for loophole-free tests of Bell inequalities 
\cite{Eberhard}, as well as allowing logical arguments that demonstrate the
nonlocality of quantum mechanics without inequalities \cite{Hardy}.
Recently, we have proposed some schemes for applications of nonmaximal
entanglement \cite{xue1,xue2,xue4}. Up to now, nonmaximally entangled states
have been deterministically generated with an ion-trap \cite{Tur}, and with
a spontaneous down converter (SPDC) \cite{Torgerson,Dig,White}.

Here, we describe an experimental scheme of preparing nonmaximally entangled
states based on Raman type laser manipulation of the atomic ensembles and
single photon detection which postselects the desired state in a
probabilistic fashion.

The basic element of this scheme is an ensemble of many identical alkali
atoms with a Raman type $\Lambda $-level configuration shown as Fig. 1, the
experimental realization of which can be either a room-temperature dilute
atomic gas \cite{Jul,Phi} or a sample of cold trapped atoms \cite{Liu,Roch}.
We continue to use the symbols and corresponding definitions in Refs. \cite
{Duan2,Duan3}. The collective atomic operator is defined as 
\begin{equation}
s=\left( 1/\sqrt{N_a}\right) \sum_{i=1}^{N_a}\left| g\right\rangle
_i\left\langle s\right| ,  \eqnum{1}
\end{equation}
where $N_a\gg 1$ is the total atom number. The Raman transition $\left|
g\right\rangle \rightarrow \left| e\right\rangle $ is coupled by the
classical laser and the forward scattered Stokes light comes from the
transition $\left| e\right\rangle \rightarrow \left| s\right\rangle $ \cite
{Duan2}. The scheme for preparation nonmaximally entangled states works in
the following way (seeing Fig. 2). Here we choose a $\Lambda $ configuration
of atomic states of $^{87}$Rb by way of example, which is coupled by a pair
of optical fields.

There are two light fields with the Rabi frequencies $\Omega $ and $\omega $%
, respectively, which couple pairs of Zeeman sublevels of electronic ground
state $5S_{1/2}$ $^{87}$Rb atoms $\left( \left| g\right\rangle ,\left|
s\right\rangle \right) $, with magnetic quantum numbers differing by two,
via the excited $5P_{1/2}$ state \cite{Phi,XP}. In this case $\left|
g\right\rangle $ and $\left| s\right\rangle $ of the simplified three-level
model correspond, respectively, to $\left| F=2,M_F=-2\right\rangle $ and $%
\left| F=2,M_F=0\right\rangle $. The atoms in the ensembles are initially
prepared to the ground state $\left| g\right\rangle $ through optical
pumping. The two ensembles 1 and 2 are illuminated by the synchronized
classical laser pulses of right circularly polarized ($\sigma _{-}$) light.
The excitations $5S_{1/2}$ $F=2$, $M_F=-2\rightarrow 5P_{1/2}$ $F=1$, $%
M_F=-1 $ can be transferred to optical excitations. Assume that the
light-atom interaction time $t_0$ is short so that the mean photon number in
the forward scattered Stokes light is much smaller than 1. It is defined in
Ref. \cite{Duan2} an effective single-mode bosonic operator $a$ for this
Stokes pulse. The whole state of the atomic collective mode and the forward
scattering Stokes mode can be written as 
\begin{equation}
\left| \varphi \right\rangle =\left| vac\right\rangle _a\left|
vac\right\rangle _p+\sqrt{p_c}s^{+}a^{+}\left| vac\right\rangle _a\left|
vac\right\rangle _p+o\left( p_c\right) ,  \eqnum{2}
\end{equation}
where $\left| vac\right\rangle _a$ and $\left| vac\right\rangle _p$ denote
the vacuum states of atomic ensembles and Stokes light respectively, and $%
p_c $ is the small excitation probability \cite{Duan2}. The forward
scattered Stokes pulses of left circularly polarized ($\sigma _{+}$) light
from both ensembles are combined at a $50\%-50\%$ beam splitter (BS) and a
single-photon detector click in one of the four detectors, after the
quarter-and half-wave plates (QWP and HWP), and polarizing beam splitter
(PBS). Adjustable QWP, HWP and PBS allow polarization analysis in any basis,
i.e., at any position on the Poincare sphere \cite{White,Born}.

Nonmaximally entangled states are produced simply by adjusting the relative
inclination between the optic axes of QWP and HWP. After some filters which
filter out the pumping laser pulses, by rotating the optic axis of QWP, the
Stokes light is turned to linearly polarized photon and the orientation of
the linear polarizer lies in the vertical plane. For an inclination $\theta
_i$ between the optic axis of HWP and the orientation of the linear
polarizer of the Stokes light in arm $i$ ($i=1,2$), the Stokes photon is $%
\sin 2\theta _i\left| H\right\rangle +e^{i\phi }\cos 2\theta _i\left|
V\right\rangle $, where $H$ and $V$, respectively, represent the horizontal
and vertical polarizations of the photon, and $\theta _1=\frac \pi 4-\theta
_2$. Then the Stokes pulses in both arms are combined at the BS and the
output light goes through a PBS, respectively, and a single-photon detector
click in one of the four detectors D1, D2, D3 and D4 measures the
combination radiation from the samples $A^{+}A$ or $A^{\prime +}A^{\prime }$%
. Here, 
\begin{equation}
A=\alpha a_1+e^{i\phi _{12}}\beta a_2,  \eqnum{3}
\end{equation}
or 
\begin{equation}
A^{\prime }=\beta a_1+e^{i\phi _{12}}\alpha a_2,  \eqnum{4}
\end{equation}
where $\phi _{12}=\phi _2-\phi _1$ is a difference of the phase shift which
is fixed by the optical channel connecting the two atomic ensembles, and $%
\alpha =\sin 2\theta _1$, $\beta =\cos 2\theta _1$. That is, if D1 or D3
clicks, the two ensembles are entangled in the form 
\begin{equation}
\left| \psi \right\rangle _{12}=\left( \alpha s_1^{+}+e^{i\phi _{12}}\beta
s_2^{+}\right) \left| vac\right\rangle _{12};  \eqnum{5}
\end{equation}
if D2 or D4 clicks, the state of the ensembles is 
\begin{equation}
\left| \psi ^{\prime }\right\rangle _{12}=\left( \beta s_1^{+}+e^{i\phi
_{12}}\alpha s_2^{+}\right) \left| vac\right\rangle _{12}\text{.}  \eqnum{6}
\end{equation}
The subscripts $1$ and $2$ are used to distinguish the atomic ensemble $E1$
and $E2$ (seeing Fig. 2). If one excitation is registered, we succeed to
entangle the two ensembles in a nonmaximally entangled state. Otherwise, we
need to repeat the above steps until we get a click in one of the four
detectors.

Now, we consider the efficiency of this scheme, which is usually described
by the total generation time. The preparation based on the Raman driving $%
\left| g\right\rangle \rightarrow \left| e\right\rangle \rightarrow \left|
s\right\rangle $, succeeds with a controllable probability $p_c$ for each
Raman driving pulse, and needs to be repeated in average $1/p_c$ times for
the final successful state generation. In the generation process, the
dominant noise is the photon loss, which includes the contributions from the
channel attenuation, the spontaneous emissions in the atomic ensembles, the
coupling inefficiency of Stokes light into and out of the channel, and the
inefficiency of the single-photon detectors which can no perfectly
distinguish between one and two photons. All the above noise is described by
an overall loss probability $\eta $. Due to the noise, the total generation
time is represented by $T\sim t_0/\left[ \left( 1-\eta \right) p_c\right] $,
where $t_0$ is the light-atom interaction time.

Also with the noise, the state of the ensembles is actually described by 
\begin{equation}
\rho _{12}=\frac 1{c+1}\left( c\left| vac\right\rangle _{12}\left\langle
vac\right| +\left| \psi \right\rangle _{12}\left\langle \psi \right| \right)
,  \eqnum{7}
\end{equation}
and 
\begin{equation}
\rho _{12}^{\prime }=\frac 1{c+1}\left( c\left| vac\right\rangle
_{12}\left\langle vac\right| +\left| \psi ^{\prime }\right\rangle
_{12}\left\langle \psi ^{\prime }\right| \right) ,  \eqnum{8}
\end{equation}
where the vacuum coefficient $c$ is basically given by the conditional
probability for the inherent mode-mismatching noise contribution \cite{Duan4}%
.

Since the nonmaximally entangled states shown in Eqs. (5) and (6) are
entangled in the Fock basis, it is experimentally hard to do certain
single-bit rotation. In the following we will show how the nonmaximally
entangled states can be used to realize the communication protocols, such as
the CHSH\ detection, with simple experimental configurations.

The first step is to share an EPR type of entangled state \cite{Duan2} 
\begin{equation}
\left| \Psi _\phi \right\rangle _{L_1R_1}=\left( s_{L_1}^{+}+e^{i\phi
}s_{R_1}^{+}\right) /\sqrt{2}\left| vac\right\rangle _{L_1R_1}  \eqnum{9}
\end{equation}
between two distant ensembles $L_1$ and $R_1$, and the presence of the noise
modifies the projected state of the ensembles to 
\begin{equation}
\rho _{L_1R_1}=\frac 1{c+1}\left( c\left| vac\right\rangle
_{L_1R_1}\left\langle vac\right| +\left| \Psi _\phi \right\rangle
_{L_1R_1}\left\langle \Psi _\phi \right| \right) .  \eqnum{10}
\end{equation}
The ensembles $L_2$ and $R_2$ are prepared in a nonmaximally entangled state 
$\rho _{L_2R_2}$ shown in Eq. (7). The $\phi $-parameters in $\rho _{L_1R_1}$
and $\rho _{L_2R_2}$ are the same provided that the two states are
established over the same stationary channels.

A basis of the ``polarization'' qubit (in analogy to the language for
photons) can be defined from the states $\left| H\right\rangle
_i=s_{i_1}^{+}\left| vac\right\rangle _{i_1i_2}$, $\left| V\right\rangle
_i=s_{i_2}^{+}\left| vac\right\rangle _{i_1i_2}$ ($i=L,R$). Single-bit
rotations in this basis can be done using the phase shift $\phi _i$ together
with the corresponding beam splitter operation with the rotation angle $%
\theta _i=\phi _i/2$ similarly to the manipulations shown in Ref. \cite
{Duan2}.

The four ensembles are illuminated by the synchronized classical laser
pulses with the frequency $\omega $. If the ensemble is in the metastable
state after the repumping pulse, the transition $\left| e\right\rangle
\rightarrow \left| s\right\rangle $ will occur {\it determinately}. We
register only the coincidences of the two-side detectors, so the protocol is
successful only if there is a click on each side. Under this condition, the
vacuum components in the entangled states and the state $%
s_{L_1}^{+}s_{L_2}^{+}\left| vac\right\rangle _{L_1L_2R_1R_2}$ and $%
s_{R_1}^{+}s_{R_2}^{+}\left| vac\right\rangle _{L_1L_2R_1R_2}$ have no
contributions to the experimental results. Then, for the measurement scheme
shown by Fig. 3, the state $\rho _{L_1R_1}\otimes \rho _{L_2R_2}$ is
effectively equivalent to the following ``polarization'' nonmaximally
entangled (PNE) state 
\begin{equation}
\left| \psi \right\rangle _{PNE}=\left( \alpha s_{L_2}^{+}s_{R_1}^{+}+\beta
s_{L_1}^{+}s_{R_2}^{+}\right) \left| vac\right\rangle _{L_1L_2R_1R_2}\text{.}
\eqnum{11}
\end{equation}
The success probability for the projection is given by $p=1/\left[ 4\left(
c+1\right) ^2\right] $.

Now, it is clear how to do the Clauser-Horne-Shimony-Holt (CHSH) inequality
detection \cite{CHSH}. We define the measurement results to be $1$ if D1 or
D3 clicks, and $-1$ if D2 or D4 clicks. Then the quantity 
\begin{equation}
E\left( \phi _L,\phi _R\right) =P_{D_1D_3}+P_{D_2D_4}-P_{D_1D_4}-P_{D_2D_3} 
\eqnum{12}
\end{equation}
is the corresponding coefficient of the measurements performed by Side L in
the basis rotated by $\phi _L/2$ and by Side R in the basis rotated by $\phi
_R/2$. According to the quantum rules 
\begin{equation}
E\left( \phi _L,\phi _R\right) =4\alpha ^2\beta ^2\cos \left( \phi _L-\phi
_R\right) ,  \eqnum{13}
\end{equation}
one can define the quantity $S$ composed of the correlation coefficients for
which both sides used analysis (phase shift $\phi _i$) of different
orientation 
\begin{eqnarray}
S &=&E\left( \phi _L^1,\phi _R^3\right) +E\left( \phi _L^1,\phi _R^2\right)
+E\left( \phi _L^2,\phi _R^3\right) -E\left( \phi _L^2,\phi _R^2\right) 
\nonumber \\
&=&8\sqrt{2}\alpha ^2\beta ^2,  \eqnum{14}
\end{eqnarray}
where $\phi _L^1=0$, $\phi _L^2=\frac \pi 2$, $\phi _L^3=\frac \pi 4$, and $%
\phi _R^1=0$, $\phi _R^2=-\frac \pi 4$, $\phi _R^3=\frac \pi 4$. Any local
realistic theory requires $S<2$. As Eq. (14) shows, $S$ varies with degree
of entanglement, for maximally entangled states, the quantity is $2\sqrt{2}$%
. For $0.479\leq \alpha \leq 0.878$ or $-0.878\leq \alpha \leq -0.479$, the
inequality is violated.

We have a brief conclusion. In this report, we describe an experimental
scheme of generating nonmaximal entanglement between two atomic ensembles.
The degree of entanglement is readily tunable. This protocol fits well the
status of the current experimental technology. Finally we use the
nonmaximally entangled states to measure the CHSH inequality.

We thank L.-M. Duan for helpful discussion, and Y.-S. Zhang and Z.-W. Zhou
for stimulating comments. This work was funded by National Fundamental
Research Program (2001CB309300), National Natural Science Foundation of
China, the Innovation funds from Chinese Academy of Sciences, and also by
the outstanding Ph. D thesis award and the CAS's talented scientist award
entitled to Luming Duan.

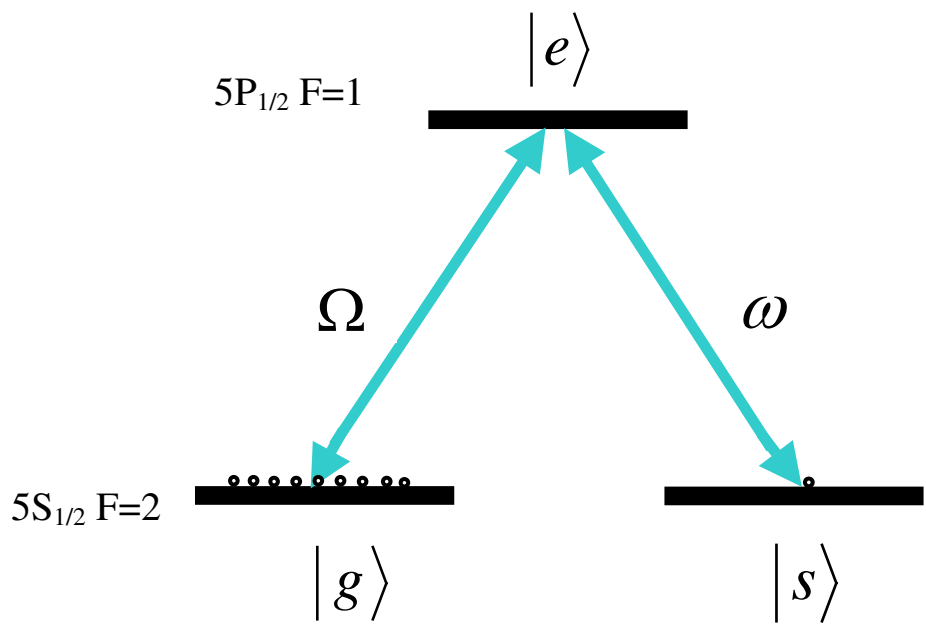
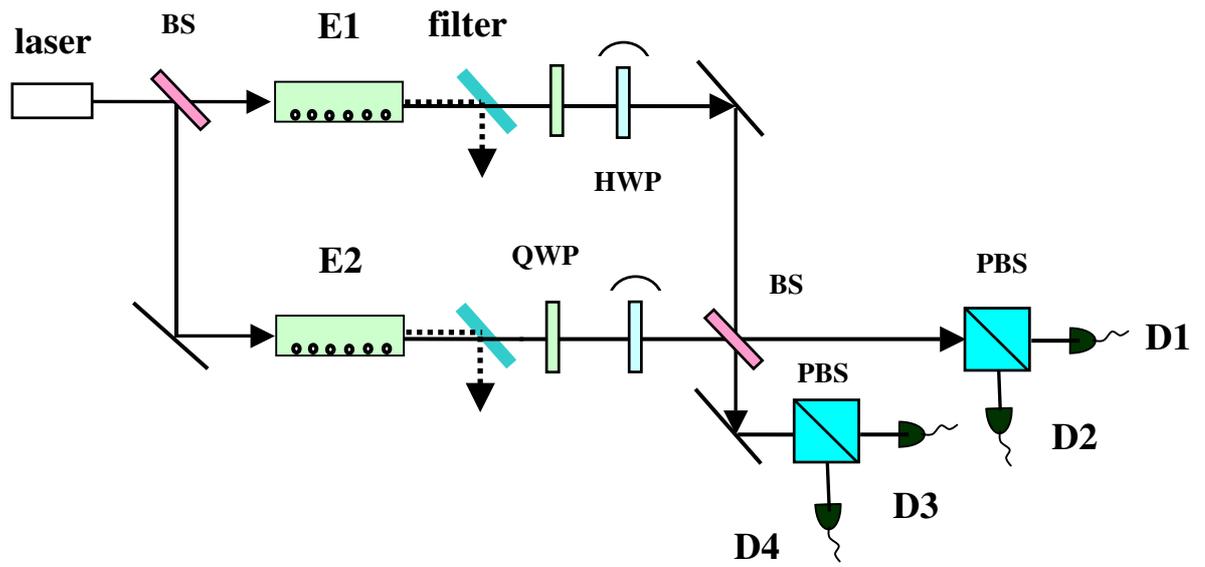
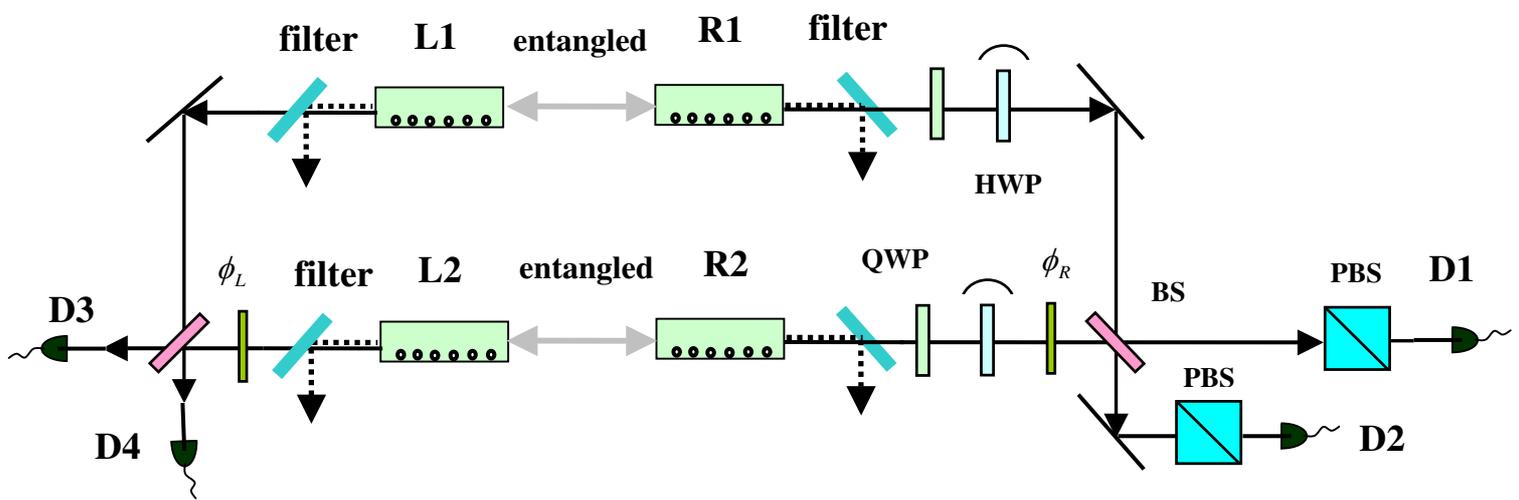
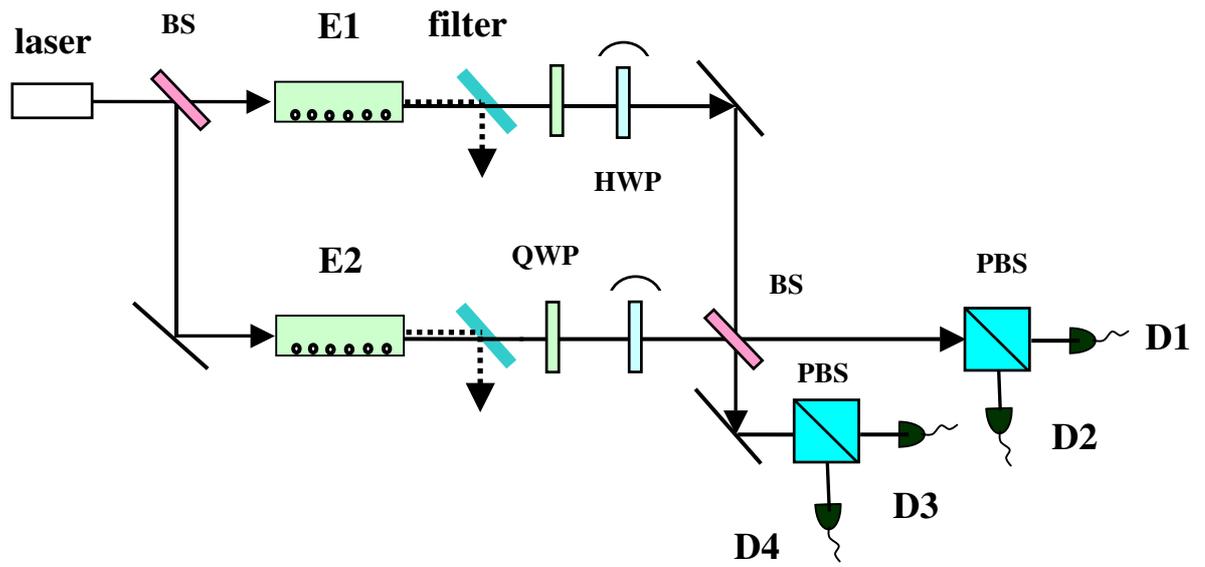
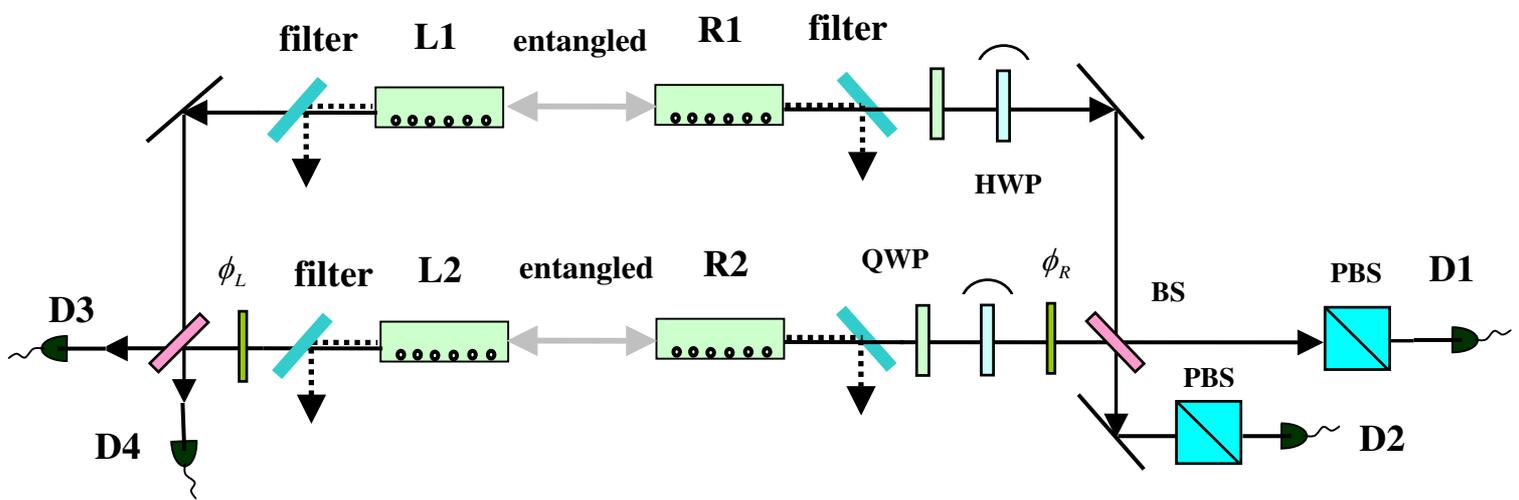
\end{document}